\documentclass{article}
\usepackage[a4paper, total={6in, 8in}]{geometry}
\usepackage{amssymb}
\usepackage{graphicx,epstopdf}
\usepackage{gensymb}
\usepackage{epsfig}
\usepackage{color}
\usepackage{epstopdf}
\usepackage{amsmath}
\usepackage{caption}
\usepackage{subcaption}
\usepackage{authblk}
\usepackage[super,comma]{natbib}
\usepackage{url}

\newcommand{\grl}{    {Geophys. Res. Lett.}}
\newcommand{\jgr}{    {J. Geophys. Res.}}
\newcommand{\ssr}{    {Space Sci. Rev.}}
\newcommand{\planss}{    {Plan. Sp. Sci.}}

\newcommand{\solphys}{ {Solar Physics}}
\newcommand{\apj}{ {Astrophys. J. }}
\newcommand{\apjl}{ {Astrophys. J. Lett.}}
\newcommand{\apjs}{ {Astrophys. J. Supplement Series}}

\newcommand{\pre}{ {Phys. Rev. E}}
\newcommand{\prl}{ {Phys. Rev. Lett.}}

\def\mP{{\cal P}} 
\def\mC{{\cal C}} 

\begin{document}


\title{Electron resonant interaction with whistler-mode waves around the Earth's bow shock I: the probabilistic approach} 

\author[1]{Xiaofei Shi}
\author[2,3]{David S. Tonoian}
\author[1,4]{Anton V. Artemyev}
\author[2,1]{Xiao-Jia Zhang}
\author[1]{Vassilis Angelopoulos}
\affil[1]{Department of Earth, Planetary, and Space Sciences, University of California, Los Angeles, USA; sxf1698@g.ucla.edu}
\affil[2]{Department of Physics, University of Texas at Dallas, Richardson, TX, USA; david.tonoian@utdallas.edu}
\affil[3]{Faculty of Physics, National Research University Higher School of Economics, Moscow, Russia, 105066.}
\affil[4]{Space Research Institute, RAS, Moscow, Russia}

\maketitle

\begin{abstract}
Adiabatic heating of solar wind electrons at the Earth's bow shock and its foreshock region produces transversely anisotropic hot electrons that, in turn, generate intense high-frequency whistler-mode waves. These waves are often detected by spacecraft as narrow-band, electromagnetic emissions in the frequency range of $[0.1,0.5]$ of the local electron gyrofrequency. Resonant interactions between these waves and electrons may cause electron acceleration and pitch-angle scattering, which can be important for creating the electron population that seeds shock drift acceleration. The high intensity and coherence of the observed whistler-mode waves prohibit the use of quasi-linear theory to describe their interaction with electrons. In this paper, we aim to develop a new theoretical approach to describe this interaction, that incorporates nonlinear resonant interactions, gradients of the background density and magnetic field, and the fine structure of the waveforms that usually consist of short, intense wave-packet trains. This is the first of two accompanying papers. It outlines a probabilistic approach to describe the wave-particle interaction. We demonstrate how the wave-packet size affects electron nonlinear resonance at the bow shock and foreshock regions, and how to evaluate electron distribution dynamics in such a system that is frequented by short, intense whistler-mode wave-packets. In the second paper, this probabilistic approach is merged with a mapping technique, which allows us to model systems containing short and long wave-packets. 
\end{abstract}

\maketitle

\section{Introduction}
The interaction between solar wind electrons and the bow shock results in compressional heating \cite{Leroy&Mangeney84,Wu84} and the formation of unstable electron populations which can generate whistler-mode waves \cite{Veltri&Zimbardo93:instability,Veltri&Zimbardo93}. A combination of the cross-field thermal anisotropy (due to compression) and a finite heat flux (typical for solar wind electrons\cite{Wilson13:waves,Tong19:ApJ}) provides free energy for whistler-mode waves in the frequency range of $\sim[0.1,0.5]$ of the local electron gyrofrequency \cite{Vasko20:pop}. Spacecraft often detect such high-frequency (high compared to whistler-mode magnetosonic waves, also observed in the bow shock\cite{Wilson16:review}), very intense (having amplitudes up to $\sim 1$\% of the background magnetic field) whistler-mode waves around the bow shock \cite{Wilson12,Hull12, Hull20, Vasko:Page:21:apjl} and in its upstream region\cite{Shi20:foreshock_whistlers}. Electron compressional heating upstream (see Ref. \cite{Liu17:foreshock&electrons}) is associated with foreshock transients \cite{Turner13:foreshock,Liu16:foreshock}, mesoscale magnetic field perturbations that are formed due to solar wind discontinuity interactions with the bow shock \cite{Lin02:hfa,Lin97:hfa, Omidi&Sibeck07,Omidi09}.

In the dense, weakly magnetized plasma around the bow shock (exhibiting a large ratio of plasma frequency to electron gyrofrequency, $f_{pe}/f_{ce}\geq 100$), high-frequency whistler-mode waves may potentially play an important role for $0.1-10$keV electron scattering \cite{Oka17,Oka19}. Such scattering is an important element of the stochastic shock drift acceleration mechanism \cite{Amano20,Morris23}. Moreover, in the strongly inhomogeneous magnetic field of the bow shock and foreshock transients, intense whistler-mode waves may provide efficient electron acceleration \cite{Kuramitsu05,Artemyev22:jgr:bowshock,Shi22:ApJ} via nonlinear resonant interactions (see reviews in Refs. \cite{Shklyar09:review,Albert13:AGU,Artemyev18:cnsns} for details). A combination of resonant scattering, shock drift acceleration (with possible contributions from nonlinear resonant interactions), and adiabatic compressional heating has been proposed to explain the formation of relativistic electrons ($\sim 100$keV, i.e., $\times 10^4$ higher than the solar wind electron temperature\cite{Wilson18:apjs}) upstream of the bow shock\cite{Wilson16:prl,Liu19:foreshock}. Explaining electron acceleration in the near-Earth space plasma environment can also be important for astrophysical systems, where shock waves are believed to play a crucial role in electron acceleration \cite{Amano22,Bohdan23}.

Although the magnetic field configuration of the bow shock and foreshock transients 
 \cite{Krasnoselskikh13,Burgess&Scholer13,Wilson16:review,Zhang22:ssr_foreshock}, as well as the statistical properties of high-frequency whistler-mode waves \cite{Hull20, Vasko:Page:21:apjl, Shi20:foreshock_whistlers,Shi22:ApJ} have been well established, the theory of electron resonant interactions with whistler-mode waves in this context is still under development. One complication is that these waves are often sufficiently intense to resonate with electrons nonlinearly (see nonlinearity criterion in Refs. \cite{Karpman74:ssr,Shapiro&Sagdeev97}). The high wave intensity invalidates the approximations of quasi-linear diffusion theory \cite{Karpman74:ssr,LeQueau&Roux87} and requires modification of standard models of wave-particle interactions (such modifications have already been made for radiation belt physics, see Refs. of \cite{Omura15,Hsieh&Omura17,Hsieh20,Vainchtein18:jgr,Artemyev18:jpp}). Another complication is that waves propagate in short wave-packets, with a small coherence length (most likely due to the modulation of the waves at their generation region \cite{Nunn86,Nunn21} and the subsequent wave trapping at the local magnetic field minima formed by compressional ultra-low-frequency perturbations \cite{Hull12, Hull20}). This requires additional modifications of models of nonlinear wave-particle interactions.

There exist two main nonlinear resonance effects: phase trapping and phase bunching \cite{Shklyar09:review,Albert13:AGU,Artemyev18:cnsns}. Both are strongly affected by the wave-packet size (i.e., the wave coherence length): the efficiency of electron acceleration via phase trapping and pitch-angle scattering via phase bunching decrease with decreasing wave-packet size because the particle does not remain in resonance with the wave for a long time\cite{Zhang18:jgr:intensewaves,Tao13,Mourenas18:jgr,Mourenas22:jgr:ELFIN}. Nonlinear resonant interactions with long wave-packets (those with $>20$ wave periods) are non-diffusive and are characterized by large energy/pitch-angle changes for a single resonant interaction \cite{Demekhov06,Omura07,Bortnik08}. As the wave-packet shortens, the interaction conforms progressively to a more diffusive treatment, as smaller energy/pitch-angle changes are attained during each resonant interaction with a wave-packet\cite{Zhang20:grl:phase,An22:Tao,Allanson20,Allanson21,Gan22}. A new modeling approach, going beyond the quasi-linear diffusion and nonlinear resonance theories' assumption of infinitely long wave-packets, is required for electron resonant interactions with whistler-mode waves around the Earth's bow shock.

\begin{figure*}
\centering
\includegraphics[trim={0 2cm 0 0},width=0.8\textwidth]{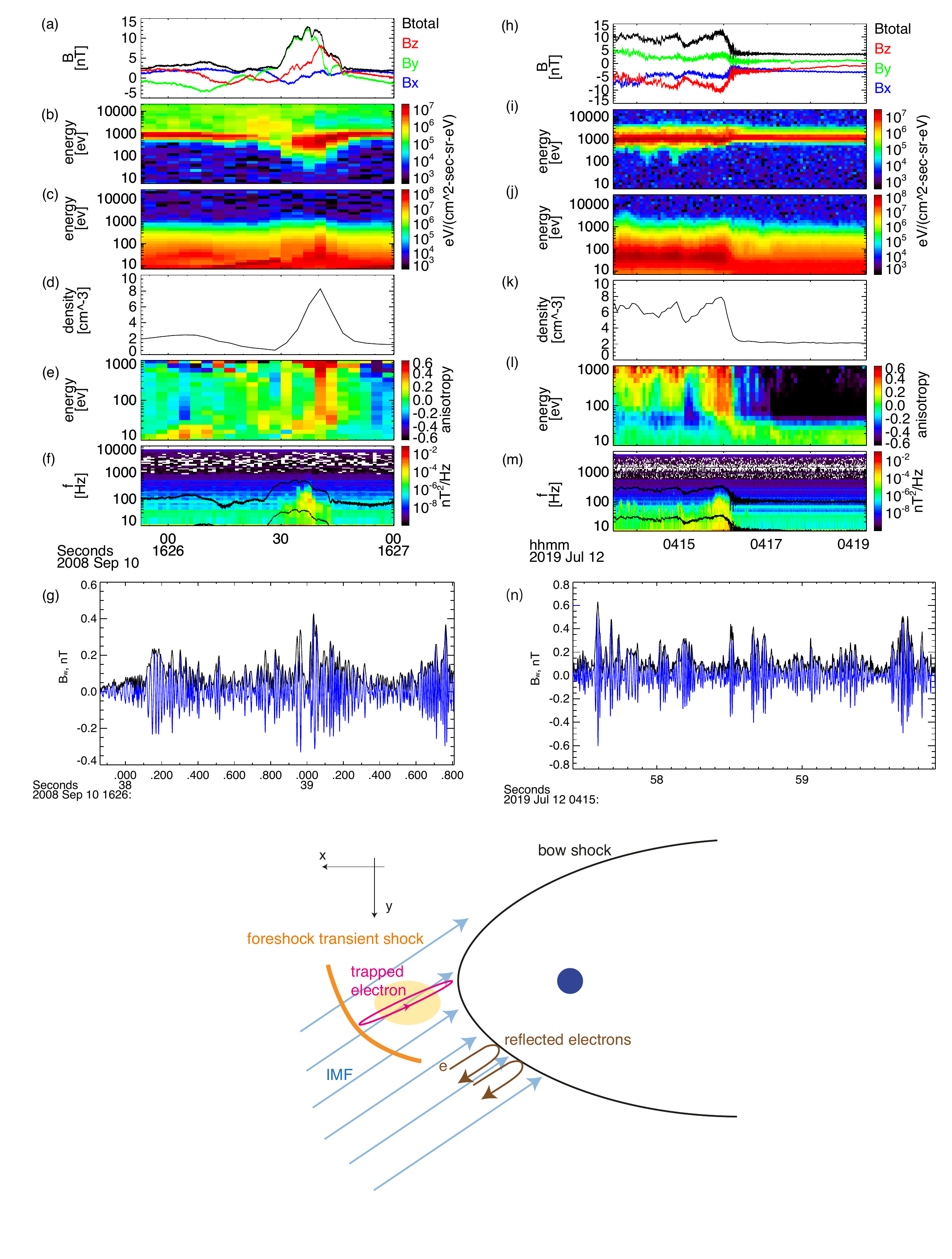}
\caption{Observations of a foreshock transient (left) and the distant bow shock (right) by the THEMIS spacecraft \cite{Angelopoulos08:ssr,Angelopoulos11:ARTEMIS}. Fluxgate magnetometer \cite{Auster08:THEMIS} measurements of magnetic fields show a strong peak in the magnetic field magnitude at the foreshock transient (a) and a step-like increase in the magnetic field magnitude at the bow shock (h). Measurements by the electrostatic analyzer\cite{McFadden08:THEMIS} show the thermalization of solar wind ion flows (b, i); an increase in the electron temperature (an increase of electron fluxes $f$ above $1$ keV) within the foreshock transient (c) and across the bow shock (j); the formation of a transversely anisotropic electron population at $>100$eV with $(f_{\perp}-f_{\parallel})/(f_{\perp}+f_{\parallel})>1$, where $f_{\perp}$, $f_{\parallel}$ are electron fluxes averaged over pitch-angle ranges of $[0,30^\circ]$ and $[75^\circ,105^\circ]$ (e,l); and density increase in correlation with the magnetic field magnitude increase (d,k). Measurements by the search coil magnetometer \cite{LeContel08,Cully08:ssr} show electromagnetic wave activities within the whistler-mode frequency range (f,m) and examples of typical whistler-mode wave-packets (g,n). Black lines in (f,m) are electron cyclotron frequency and $1/10$ of this frequency. Black and blue lines in (f,m) are wave magnetic field magnitude and one of the two transverse magnetic field components. 
The bottom panel depicts the electron dynamics (brown: reflection; purple: trapping) and wave-particle resonant interactions (in a yellow-highlighted interaction region) at the bow shock (black) and within the foreshock transient (demarcated by the bow shock and the shock upstream of it in orange).
\label{fig1} } 
\end{figure*}

We have developed such a probabilistic approach model based on the one proposed for wave resonant interactions within electrons bouncing along magnetic field lines in the radiation belts \cite{Tao08:stochastic,Artemyev17:pre,Lukin21:pop}. The approach assumes multiple, independent resonant interactions described by probability distribution functions (constructed theoretically or derived numerically) of energy and pitch-angle changes during each interaction. Such multiple resonant interactions are possible when particles follow the bounce motion (and periodically attain resonance with the waves) or when they resonate with multiple waves (different wave-packets spatially distributed along particle trajectories). Figure \ref{fig1} (bottom) depicts the electron dynamics and wave-particle interaction for a foreshock transient (an example is shown on the left) and for a bow shock (an example is shown on the right). In both environments, the background plasma density varies strongly with the background magnetic field (see Fig. \ref{fig1}(a,d) and (h,k)), and the $f_{pe}/f_{ce}$ ratio is large ($\approx 100$) and almost constant. The whistler-mode wave characteristics are also quite similar in these two environments: waves propagate in the form of intense ($\sim 1$\% of background magnetic field), short wave-packets (see Fig. \ref{fig1}(g,n)). This allows us to apply our model to both systems, with equal efficacy. For the bow shock region, we are interested in reflected electrons, which can be scattered by whistler-mode waves upstream and turned back to the shock. These electrons should resonate with waves that are generated upstream and propagating downstream. For the foreshock transients, electrons may be trapped between the shock of the foreshock transient and the bow shock. In that case, the electrons bounce back and forth, undergoing multiple resonant interactions with waves generated within the core of the foreshock transients.

This is the first of two accompanying papers. In this paper, we describe the basic properties of electron resonant interactions with whistler-mode waves and develop a probabilistic approach, which enables tracing the long-term evolution of electron distribution functions. This approach assumes that waves are sufficiently incoherent (wave-packets are short) to reduce the efficiency of nonlinear resonant interactions and hence lead to diffusive particle scatterings. Such a diffusion by intense wave-packets is quite different from the quasi-linear diffusion \cite{Shklyar21,Frantsuzov23:jpp,Gan22}. In the second paper, we describe how this probabilistic approach can be merged with a mapping technique\cite{Benkadda96,Khazanov13,Khazanov14,Artemyev20:pop} to model electron dynamics in systems with a significant effect from nonlinear resonance with long wave-packets. 

The rest of the paper includes the following sections: descriptions of the basic equations for wave-particle resonant interactions (Sect. \ref{sec:equation}), descriptions of the probability distribution of energy and pitch-angle changes (Sect. \ref{sec:pdf}), a numerical verification of the probabilistic approach (Sect. \ref{sec:verification}), and discussion of the results (Sect. \ref{sec:discussion}). In Sections \ref{sec:equation} and \ref{sec:pdf}, we examine the effect of wave-packet size on nonlinear resonance and determine the parametric range of diffusive electron scattering by intense, short wave-packets. Most examples and conclusions are valid for electron dynamics in both bow shock and foreshock transients. 

\section{Basic equations}\label{sec:equation}
We use the Hamiltonian model for the nonrelativistic electron (mass $m_e$, charge $-e$) dynamics \cite{Artemyev22:jgr:bowshock}
\begin{equation}
H = \frac{{p_\parallel^2 }}{{2m_e }} + \mu \Omega _0 (s) - e\Phi \left( s \right) \label{eq1}
\end{equation}
where $(s,p_\parallel)$ are conjugate field-aligned coordinates and momentum, the electron cyclotron frequency $\Omega_0(s)=eB_0(s)/m_ec$ is determined by the ambient magnetic field profile, $B_0(s)$, and $\Phi(s)$ is the electrostatic potential due to decoupling of ion and electron motions around magnetic field gradients \cite{Goodrich&Scudder84,Scudder95,Gedalin96}. This model omits the effect of electron heating due to interaction with shock waves, i.e., the 
Hamiltonian equation is written in the bow shock (or foreshock transient) reference frame, but there is no projection of the shock speed on the field-aligned direction, $v_D$. The effect of this shock motion can be added as $s\to s+v_Dt$ (see Refs. \cite{Leroy&Mangeney84,Wu84}), which does not influence the wave-particle resonant interaction, because $v_D\leq 1000$km/s (see Ref. \cite{Wu84}) is smaller than the typical resonant electron velocity $\geq 5000$km/s (see Refs. \cite{Shi22:ApJ,Artemyev22:jgr:bowshock}). 

To describe the resonant interaction of electrons and field-aligned whistler-mode waves, we add the term $U_w\cos(\phi+\psi)$ to the Hamiltonian (\ref{eq1}), see Refs. \cite{Shklyar09:review,Albert13:AGU,Artemyev15:pop:probability}. The wave phase $\phi$ is determined by the wave number $k(s)=\partial\phi/\partial s$ and the wave frequency $\omega=-\partial\phi/\partial t$, whereas $k=(\Omega_{pe}/c)\cdot(\Omega_0/\omega-1)^{-1/2}$ is given by the cold plasma dispersion relation\cite{bookStix62}. The electron gyrophase $\psi$ is conjugate to the magnetic moment $\mu=E\sin^2\alpha/\Omega_0(s)$, where $E$ is the electron energy and $\alpha$ is the electron pitch-angle. The effective wave energy is $U_w=\sqrt{2\mu\Omega_0/m_ec^2}eB_w/k$, where $B_w$ is the wave magnetic field amplitude \cite{Albert93,Vainchtein18:jgr}. The system under consideration contains two smallness parameters, $\max B_w/\min B_0\ll1 $ and $1/\min k L \sim \max B_w/\min B_0 \ll 1$, where $L$ is the spatial scale of the ambient magnetic field variation, $\Omega_0=\Omega_0(s/L)$. The first parameter determines whether the wave term $\sim U_w$ is a small perturbation in the Hamiltonian $H$. The second determines whether the wave phase and gyrophase change much faster than the particle field-aligned coordinate $s$ and momentum $p_\parallel$. The resonance condition for this system is $\dot\phi+\dot\psi=0$:  
\[
k\left( s \right)\frac{{\partial H}}{{\partial p_\parallel }} - \omega  + \frac{{\partial H}}{{\partial \mu }} = k\left( s \right)\frac{{p_\parallel }}{{m_e }} - \omega  + \Omega _0 (s) = 0
\]

We build two models for two systems: Earth's bow shock and foreshock transients. For the bow shock, the minimum of the background magnetic field is far upstream, $s\to-\infty$, and the magnetic field is modeled to increase as $\Omega_0=\Omega_{\min}\cdot\left(1+3b(s)\right)$ and $b(s)=(1/2)\cdot\left(1+\tanh(s/L)\right)$. The spatial scale $L$ is about $1000$km, i.e., $1/\min k L\approx c/L\min\Omega_{pe}\sim 10^4$. This system also has a cross-shock electrostatic potential $\Phi(s)=\Phi_0 b(s)$, and plasma frequency $\Omega_{pe}/\Omega_{pe,\min}=100\cdot\sqrt{\Omega_{0}(s)/\Omega_{\min}}$. We consider electrons that move toward the shock, then are reflected by the shock, and resonate with whistler-mode waves propagating toward the shock in the upstream region. 

The foreshock transient model is characterized by a local magnetic field minimum $\Omega_0=\Omega_{\min}\cdot\sqrt{1+(s/L)^2}$, and we set $\Phi=0$. For foreshock transients, the plasma frequency $\Omega_{pe}=100\cdot\Omega_{0}(s)$ (see observations in Ref. \cite{Shi22:ApJ}). The spatial scale $L$ is also about $1000$km. We consider waves that are generated in the core region (minimum magnetic field) and propagating to both sides. As a result, electrons bouncing within the {\it magnetic bottle} forming between the bow shock and the foreshock transient shock can resonate with such waves twice per bounce period. 

\begin{figure}
\centering
\includegraphics[width=0.5\textwidth]{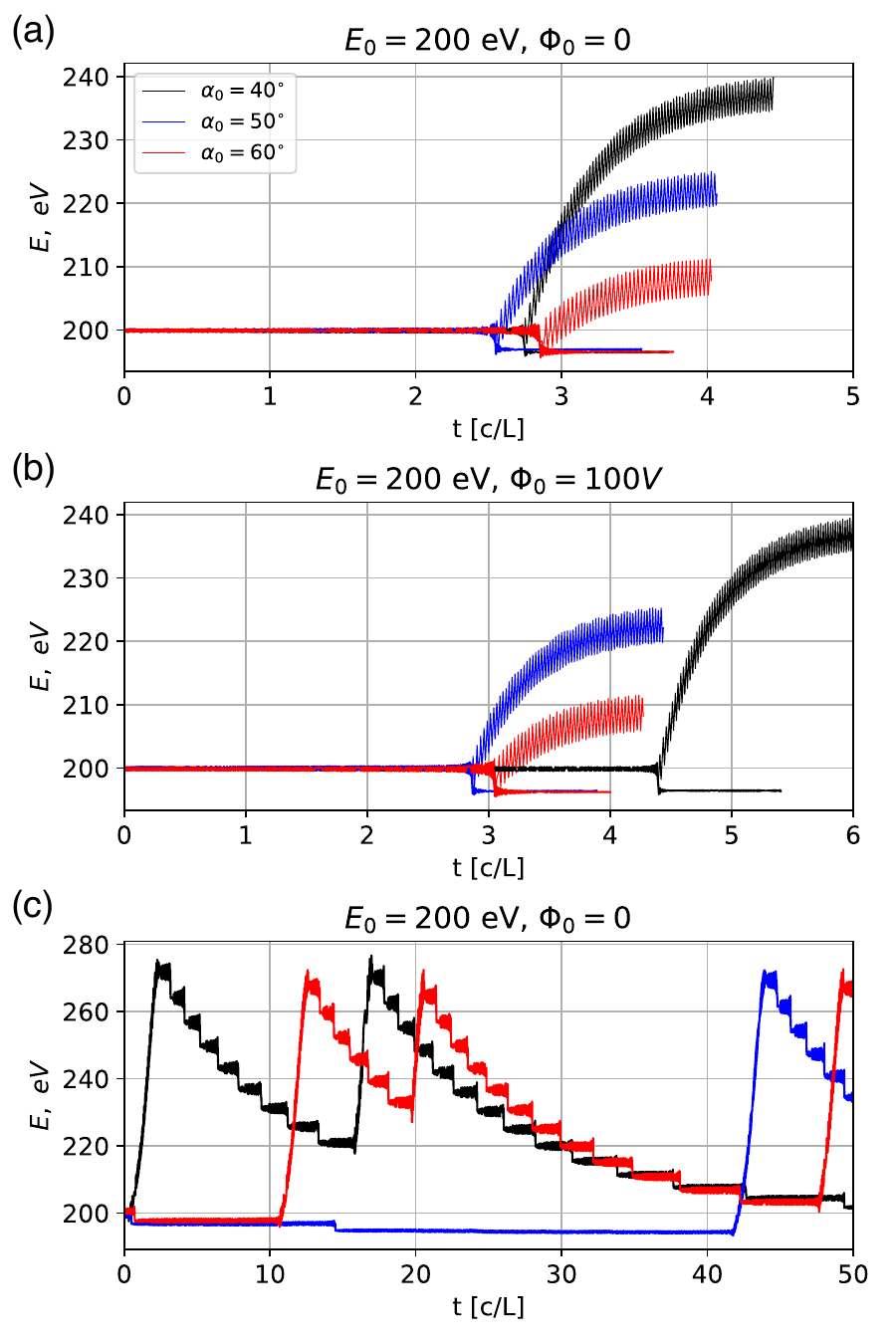}
\caption{Examples of electron energy evolution due to the interactions with the prescribed wave field  (energy versus time): for the bow shock model with $\Phi_0=0$ (a), for the bow shock with $\Phi_0=100$V (b), and for the foreshock transient (c). Each panel shows three trajectories with the same initial energy $200$eV and three different initial pitch-angles.
\label{fig2} } 
\end{figure}

Figure \ref{fig2} shows examples of electron trajectories from the numerical integration of the Hamiltonian equations of motion for two systems with different parameters. There are two main effects of nonlinear resonant interactions: phase bunching characterized by energy decrease and phase trapping characterized by large energy increase (see Refs. \cite{Karpman75PS,Trakhtengerts03,Albert00,Omura91:review}). We aim to describe electron distribution dynamics driven by multiple nonlinear resonant interactions.

\section{Probability distribution function of $\Delta E$}\label{sec:pdf}
To characterize wave-particle resonant interactions, we will use the probability distribution $\mP(\Delta E)$ of energy change for a single resonant interaction. For fixed system parameters, this distribution will depend on the initial electron energy $E$ and pitch-angle $\alpha_0$. The pitch-angle can be substituted by the initial electron magnetic moment $\mu_0=E\sin^2\alpha_0/\Omega_{\min}$ (we set initial conditions at the magnetic field minimum). Equation (\ref{eq1}) after adding the wave term, $U_w\cos\left(\phi+\psi\right)$, has one integral of motion: $E-\omega\mu=h=const$ (see, e.g., Ref. \cite{Shklyar09:review}). This is a constant because there are no electric fields in the reference frame moving with the wave ($t\to t+\omega t$), and the particle energy $\sim h$ in this reference frame is conserved. Therefore, we can use the probability distribution function $\mP(\Delta E, E)$ defined in the $(\Delta E, E)$ space for fixed $h$. For waves with infinitely long wave-packets $\sim \cos\left(\phi+\psi\right)$, such probability distribution functions can be derived analytically \cite{Vainchtein18:jgr,Artemyev22:jgr:Landau&ELFIN}. In reality, however, waves propagate in the form of wave-packets (see Fig. \ref{fig1} and Refs. \cite{Hull12,Hull20,Shi22:ApJ}), where $\mP$ can only be determined from numerical simulations \cite{Omura15,Hsieh&Omura17,Artemyev19:cnsns}. We introduce the wave field modulation, $\cos\left(\phi+\psi\right)\to f(\phi)\cos\left(\phi+\psi\right)$, and numerically evaluate $\mP$ for different modulation characteristics. Function $f(\phi)$ describes the wave-packet train, and we use a simple form $f(\phi)=\exp\left(-5\cdot\cos^2(\phi/2\pi\beta)\right)$ with $\beta$ denoting the wave-packet size \cite{Tsai22}. An additional model parameter is the wave phase coherence number measured by the number of contiguous coherent wave packets, $N_c$. This parameter describes how many wave packets within the train have the same initial $\phi$, i.e., maintain phase coherence from one packet to the next one. The limit $N_c\to \infty$ corresponds to the situation when all wave packets are generated in the same source region by the same particle population, so there is no variation (destruction) of the wave phase among wave packets (in this case electrons may be trapped into the next wave packet after escaping from the previous one, and such multi-trapping would result in effective electron acceleration, similar to infinitely long wave-packets\cite{Hiraga&Omura20}). However, different wave-packets are often generated in different source regions and their phases are not coherent across the entire packet train \cite{Zhang20:grl:phase}. This effect can be modeled by a finite $N_c$, which will reduce the efficiency of the phase trapping and acceleration.  

\subsection{Long wave-packets}
\begin{figure}
\centering
\includegraphics[width=0.35\textwidth]{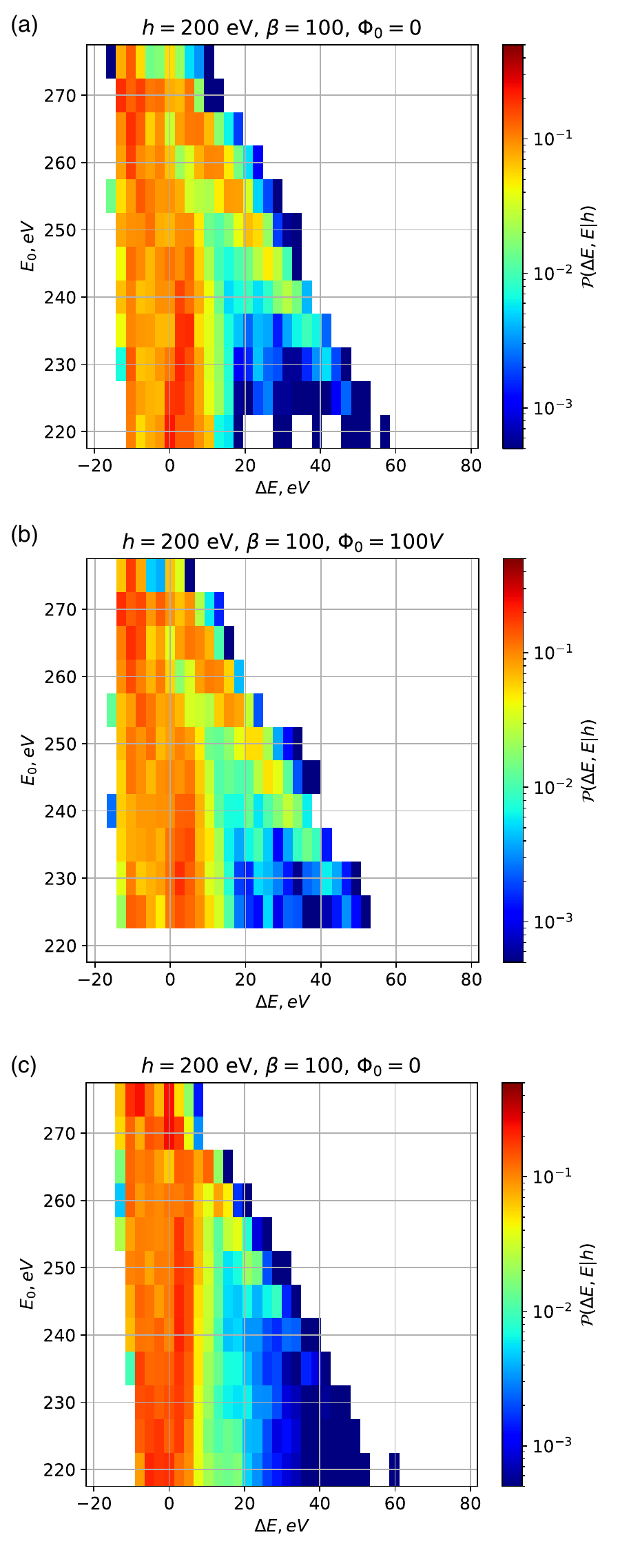}
\caption{Examples of $\mP(\Delta E, E_0)$ distributions for the bow shock model with $\Phi_0=0$ (a), for the bow shock model with $\Phi_0=100$V (b), for the foreshock model (c); in all cases, $N_c\to\infty$ and $\beta=100$. The main difference between (a) and (b) is that for $\Phi_0=100$V many electrons with resonant energies $\leq 225$eV cross the bow shock and do not resonate with waves in the upstream region.
\label{fig3} } 
\end{figure}

During a simulation of many electrons interacting with wave-packets of a given frequency, $N_c$ and $\beta$, the value of $h$ remains fixed throughout the resonant interaction, and the resultant energy change can be obtained from a probability distribution of $\Delta E$ for the fixed $h$. We can therefore use a $\Delta E$ lookup table for the given fixed $h$ of an electron of an initial pitch angle and associated resonance energy. The full range of initial equatorial pitch angles ($30^{\circ}$ to approximately $80^{\circ}$ at the minimum of the magnetic field magnitude) will map to a range of resonance energies around $220-280$eV and result in a 2-D probability distribution quantifying the results of the interaction. Figure \ref{fig3} shows probability distributions $\mP(\Delta E, E)$ for the fixed $h$ and long wave-packets ($\beta=100$, $N_c\to \infty$). There is a clear dependence of $\Delta E$-distribution on initial energy, $E_0$. For small $E_0$, the distribution of energy changes shows two distinct populations: a small number of electrons with very large $\Delta E>0$ are electrons accelerated via phase trapping, whereas the main electron population has small $\Delta E<0$ due to phase bunching. With the increase of the initial energy, $E_0$, the trapping acceleration becomes less effective and the trapped population moves closer to $\Delta E\sim 0$. This is caused by the value of the resonance energy: for fixed $h$, smaller $E_0$ means smaller $\alpha_{0}$ and larger $s$ for the resonant interactions where electrons will be trapped. As all trapped electrons escape from the resonance at $s\sim 0$, which corresponds to minimum $B_0$, the duration of electron trapping increases with larger values of $s$ (see Refs. \cite{Shklyar09:review,Omura15,Artemyev15:pop:probability} for a discussion of trapped electron acceleration in an inhomogeneous magnetic field). This leads to longer trapping times and more significant acceleration. Therefore, for long, coherent wave-packets (large $\beta$, large $N_c$) the probability distribution of energy changes, $\mP(\Delta E,E)$, depends on two parameters $(E, h)$. To describe electron dynamics, we need to determine the probability distribution in 3D space of $(\Delta E, E_0, h)$. This can be done analytically because $\Delta E$ can be determined from analysis of the Hamiltonian equation (\ref{eq1}) with the wave term included \cite{Vainchtein18:jgr,Artemyev20:pop}, and we will provide such a solution in the second paper.   

However, as we shall see in the next subsection, a further simplification in the statistical description of the resonant interactions is possible when the wave-packets are short. In that case, the dependence on $E_0$ is weak, and the interactions can be well-described as a 1-D probability distribution over a narrower energy range. This simplification with the help of a cumulative probability function leads to the probabilistic approach which we will discuss in the next section.

\subsection{Short wave-packets}
\begin{figure}
\centering
\includegraphics[width=0.35\textwidth]{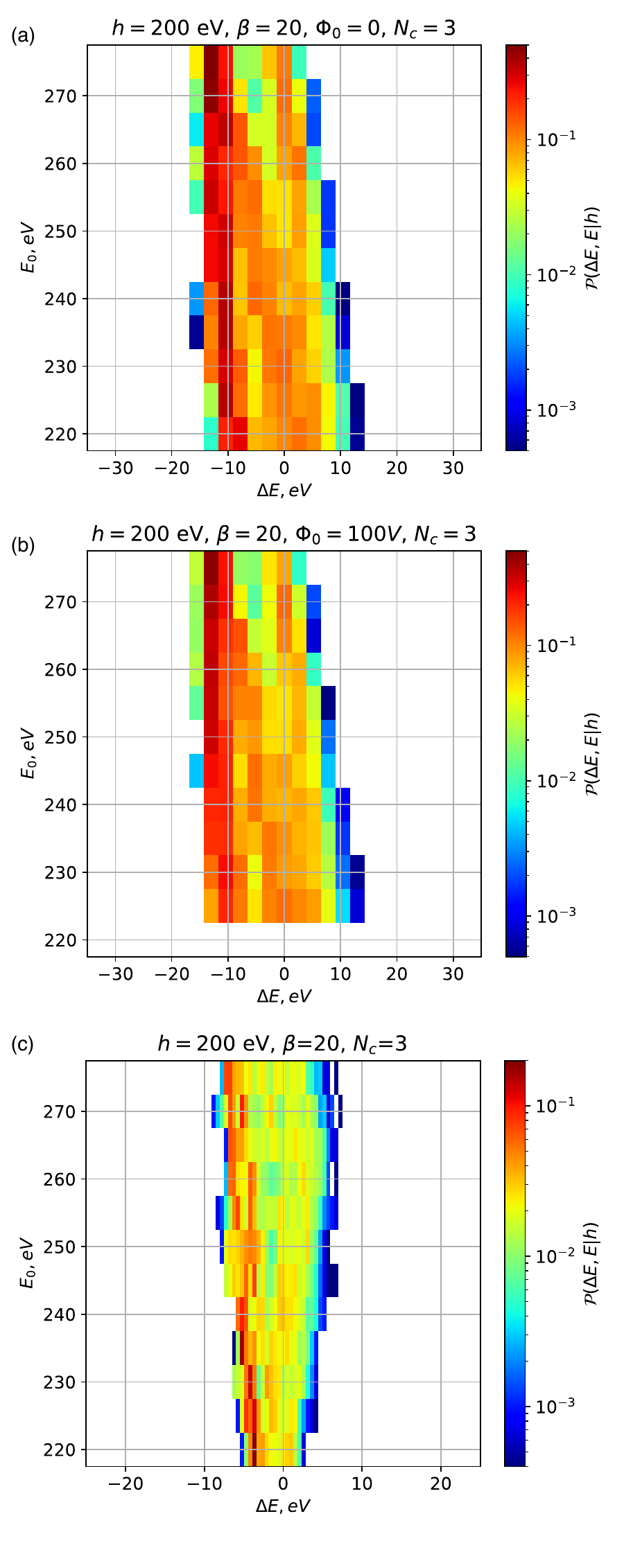}
\caption{Examples of $\mP(\Delta E, E)$ distributions for the bow shock model with $\Phi_0=0$ (a), for the bow shock model with $\Phi_0=100$V (b), for the foreshock model (c); $N_c=10$ and $\beta=20$. The main difference between (a) and (b) is that for $\Phi_0=100$V a considerable number of electrons with resonant energies $\leq 225$eV cross the bow shock and do not resonate with waves in the upstream region.
\label{fig4} } 
\end{figure}

Although it is possible to obtain an analytical model for resonant interactions between long wave packets and electrons, such long wave packets are rarely observed. In contrast, the majority of observations are of short wave packets (see Fig. \ref{fig1} and Ref. \cite{Shi22:ApJ}). For these, the efficiency of trapping acceleration varies from one interaction to another, and the role of initial conditions becomes much less important. To illustrate this, we calculate the $\mP(\Delta E, E)$ distribution for fixed $h$ and small $\beta$, $N_c$. Figure \ref{fig4} shows that in the short wave-packet limit, the $\Delta E$ -distribution is mainly distributed within $20$eV of $\Delta E$ = 0, with no significant probability of large positive values of $\Delta E$. This implies that the energy gained by trapped electrons decreases(due to the shorter time that electrons spend in the trapping acceleration, see, e.g., Refs \cite{Tao13,Zhang18:jgr:intensewaves,Mourenas18:jgr}), while the number of trapped electrons increases (due to large gradient of the wave amplitude at the edge of wave-packet, see, e.g., Ref. \cite{Bortnik08,Artemyev19:cnsns,An22:Tao} and the second paper). Such a $\Delta E$-distribution can be characterized by two parameters, $\langle \Delta E\rangle$ and $\langle (\Delta E)^2\rangle$, i.e., the wave-particle interaction is diffusive. However, this diffusion, $\langle (\Delta E)^2\rangle\propto B_w^\kappa$ with $\kappa\sim 1$, caused by almost monochromatic intense waves is quite different from the quasi-linear diffusion with $\langle (\Delta E)^2\rangle\propto B_w^2$ (see the theoretical model for $\langle (\Delta E)^2\rangle$ in Ref. \cite{Frantsuzov23:jpp}). In addition to the symmetric distribution of $\Delta E$, short wave-packets also have another important effect. As depicted in Figure \ref{fig4}, $\mP(\Delta E, E)$ now exhibits weak dependence on $E$ for fixed $h$. Therefore, instead of a 2D $\mP(\Delta E, E)$, we can use a 1D distribution, $\bar\mP(\Delta E)=\langle \mP(\Delta E, E) \rangle_E$, to describe the interaction around the specific resonance energy $E$,  for a fixed $h$. 

\section{Verification of the probabilistic approach}\label{sec:verification}
\begin{figure}
\centering
\includegraphics[width=0.5\textwidth]{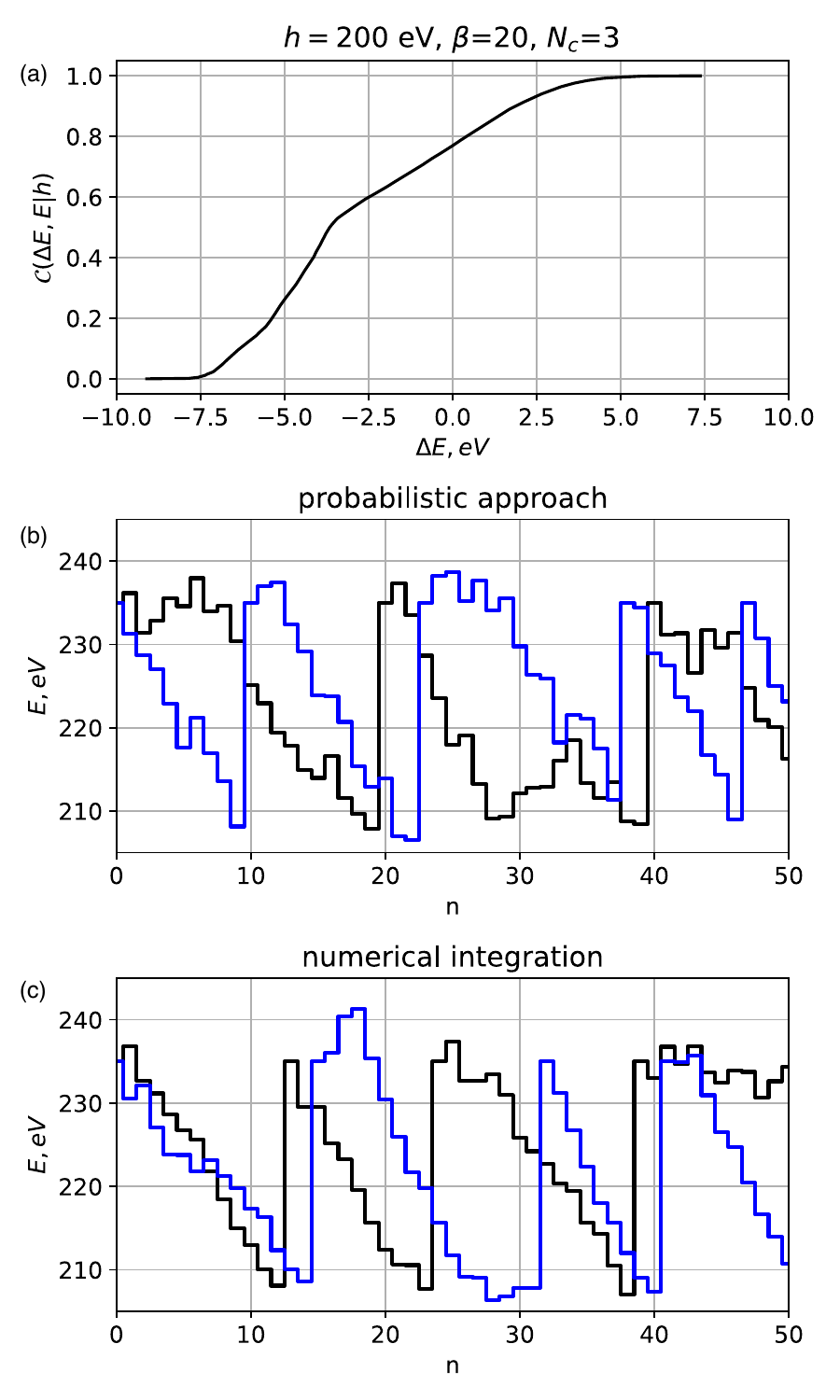}
\caption{Panel (a) shows the cumulative probability distribution function $\mC(\Delta E)$ for $\bar\mP(\Delta E)$ from Fig. \ref{fig4}(c). Panels (b,c) shows electron energy as a function of the number of resonant interactions $n$ (two interactions per bounce period) for several electron trajectories evaluated with the probabilistic approach (b) and with numerical integration of original Hamiltonian equations (c). The system configuration and parameters are the same as in Fig. \ref{fig4}(c).
\label{fig5} } 
\end{figure}

Figure \ref{fig4} shows that for systems with short wave packets, we may use a 1D $\bar\mP(\Delta E)$ distribution. This distribution is equivalent to the cumulative probability distribution $\mC(\Delta E)=\int_{-\infty}^{\Delta E}{\bar\mP(x)dx}$, which can be used for tracing the electron resonant energy change: given a random number $\xi_n \in [0,1]$, one can find the corresponding $\Delta E$  using $\mC(\Delta E)=\xi_n$ distribution. The energy change for each interaction is
$E_{n + 1}  = E_n  + \Delta E\left( {\xi _n } \right)$, where $n$ is the iteration number (number of resonant interactions). Figure \ref{fig5}(a) shows $\mC(\Delta E)$ for  $\bar\mP(\Delta E)=\langle\mP(\Delta E, E)\rangle_E$ from Fig. \ref{fig4}(c), whereas Fig. \ref{fig5}(b) shows several trajectories $E_n$ evaluated with this probabilistic approach. For comparison, we also plot $E_n$ trajectories (Fig. \ref{fig5}(c)) obtained from the numerical integration of original Hamiltonian equations with the system parameters in the caption of Fig. \ref{fig4}. Figures \ref{fig5}(b,c) demonstrate that this $\mC$-based probabilistic approach and the numerical integration approach provide very similar patterns of electron energy dynamics.

We use the foreshock transient model because for this situation we may reproduce multiple resonant interactions with the wave having a fixed frequency (also fixed $h$) as electrons bounce between mirror points. Resonant wave-particle interactions alter electron pitch-angles and thus may move their mirror points sufficiently far from $s=0$ to cause electron losses. We set the magnetic field trapping ratio as $\max B/\min B=10$, in agreement with statistical observations of magnetic field variations in the foreshock transients \cite{Shi22:ApJ}. Thus, electrons with pitch-angles (referenced at magnetic field minimum) below $\sim 18^\circ$ escape from the foreshock transient. The energy range of resonant interactions for fixed $h$ (e.g., as shown for $\mP(\Delta E, E_0)$ in Figs. \ref{fig4}, \ref{fig5}) includes such small pitch-angles (corresponding to $h=E-\omega\mu \approx E$). It is important to note two significant points regarding small pitch-angle/field-aligned electrons. First, field-aligned electrons can escape from the foreshock transients and cross the bow shock without reflection. Second, field-aligned electrons resonate with waves in a specific regime that excludes pitch-angle/energy reduction \cite{Lundin&Shkliar77}. The nonlinear resonant interactions in this regime are characterized by $\sim 100$\% phase trapping with pitch-angle/energy increase (see results of numerical simulations in Ref. \cite{Kitahara&Katoh19,Gan20:grl} and theoretical models in Refs. \cite{Albert21,Artemyev21:pop}).  These two effects complicate the modeling of field-aligned electron dynamics. To resolve this issue, we employ the following procedure: electrons with pitch-angles below $18^\circ$ are allowed to escape from the foreshock transient; however, to keep the number of electrons unchanged we re-introduce the same number of electrons into the system with their initial pitch-angles and energies. The replacement electrons can be thought of as solar wind electrons that become trapped within the foreshock transient.

\begin{figure}
\centering
\includegraphics[width=0.5\textwidth]{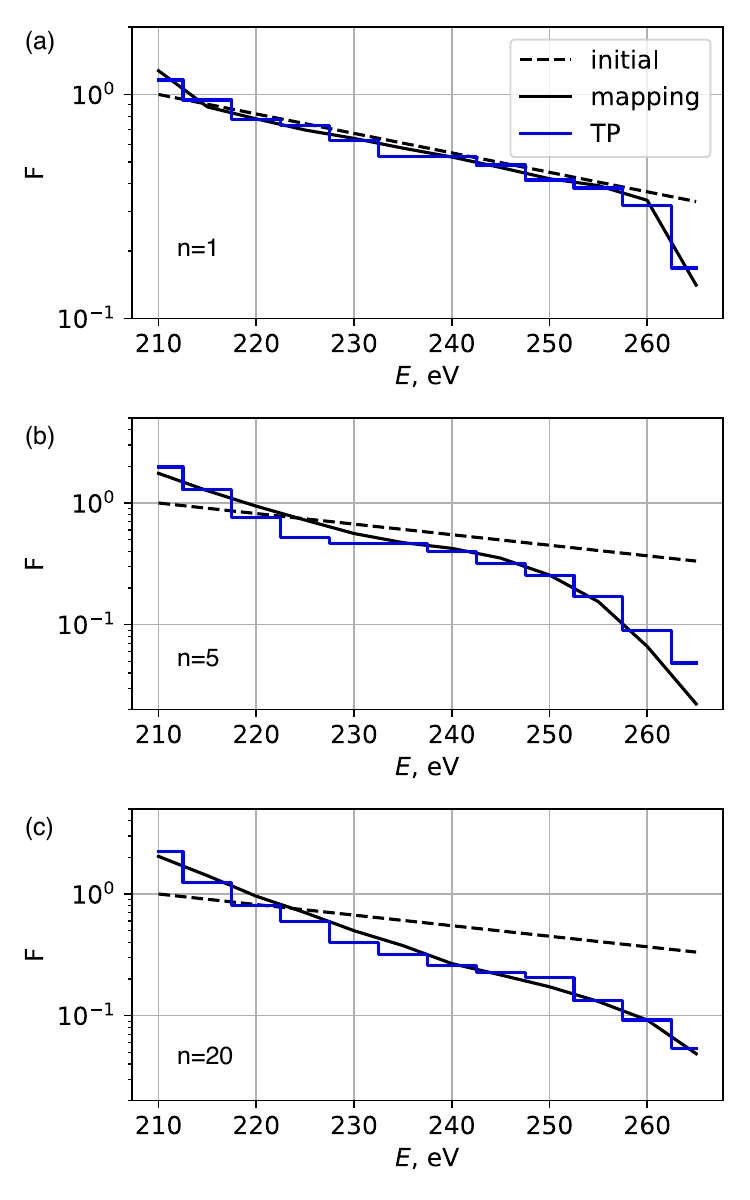}
\caption{Evolution of electron distribution function evaluated by numerical integration of Hamiltonian equations (\ref{eq1}) and by the probabilistic approach using the cumulative distribution $\mC$ of Fig. \ref{fig5}(a). Three panels show results for $n=1$(a), $n=5$(b), and $n=20$(c); the dashed curve in each panel shows the normalized initial distribution, $F(E)\sim\exp(-E/50{\rm eV})$; blue curves show results of test particle simulations, black curves show results of the probabilistic approach. \label{fig6} } 
\end{figure}

To further verify the probabilistic approach, we use the $\mC$ function to evaluate energies of $10^5$ trajectories for $20$ resonant interactions. We use the initial distribution function $F(E)\sim\exp(-E/50{\rm eV})$ to set the initial phase space density $F(E)$ and then trace its evolution in energy. Note that this is a 1D distribution for a fixed $h$; to trace the dynamics of a 2D (energy, pitch-angle) distribution, numerous $h$ should be used\cite{Vainchtein18:jgr}. To compare and validate the results obtained from the probabilistic approach, we use $10^4$ numerically integrated electron trajectories with the same initial distribution $F(E)$. Figure \ref{fig6} shows $F_m (E)$ for the probabilistic approach and $F_t(E)$ for test particle simulations with different $n$. Note that although we use iteration numbers instead of time, the same approach can be used for the time iteration, $t_{n+1}=t_n+\tau(E_n)/2$,  with $\tau(E_n)$ being the bounce period of electrons, i.e., $\tau(E_n)/2$ is the time between two resonant interactions. The scattering caused by short wave-packets mostly results in an energy decrease (which also means the pitch-angle decrease), and $F_t(E)$, $F_m(E)$ grow at smaller energies. The electron distribution obtained by test particle simulations evolves very similar to that obtained from the probability approach, i.e., the probability approach works well.  

\section{Discussion} \label{sec:discussion}
This is the first of two papers devoted to developing a theoretical approach for modeling solar wind electron scattering and acceleration by high-frequency whistler-mode waves around the Earth's bow shock. We examine the resonance effects of the most widespread wave population which consists of short, intense wave-packets. Although an individual resonant interaction with a short wave packet includes nonlinear effects such as phase trapping and bunching \cite{Omura91:review,Shklyar09:review,Albert13:AGU}, the magnitude of energy/pitch-angle change due to trapping and bunching are comparable, and the particle dynamics resembles diffusion \cite{Artemyev21:pre}. The corresponding diffusion coefficients, however, differ from the coefficients predicted by the quasi-linear diffusion theory \cite{Frantsuzov23:jpp}. Thus, instead of using linear theory and the Fokker-Plank equation \cite{bookSchulz&anzerotti74,bookLyons&Williams}, we use the probabilistic approach. This approach repeats the general stochastic differential equation method for wave-particle resonant interactions \cite{Tao08:stochastic}, but does not make any assumptions about the diffusive nature of the interactions \cite{Artemyev19:cnsns,Lukin21:pop}. 

Interactions between the solar wind electrons and the bow shock/foreshock transients include multiple effects operating on different temporal and spatial scales: electron acceleration by ion-scale electrostatic fields that form due to decoupling of ion and electron motions around strong magnetic field gradients \cite{Leroy&Mangeney84,Wu84,Scudder95}; electron demagnetization and scattering by small-scale, intense electrostatic fields \cite{Balikhin93,See13,Gedalin20}; electron trapping and adiabatic pumping by ultra-low frequency compressional fluctuations of the magnetic field \cite{Lichko&Egedal20}; electron spatial mixing by turbulent electromagnetic fields \cite{Mitchell&Schwartz13,Mitchell&Schwartz14}, etc. The description of most of these effects requires temporal and spatial resolution on the scales of ion kinetics, which is reachable in modern hybrid simulations \cite{Omidi09,Omidi16,vonAlfthan14,Lin14:hybrid_code,Omelchenko21}. The test particle approach for electron tracing in electromagnetic fields from hybrid simulations can describe all details of hot electron dynamics \cite{Liu19:foreshock} except wave-particle interaction effects, which are principally important for electron scattering and shock drift acceleration\cite{Amano22}. Thus, in the future we propose to incorporate resonant wave-particle interaction effects in test particle simulations by employing the probabilistic approach (such an approach has been extensively used at the Earth’s inner magnetosphere \cite{Elkington18:agu,Elkington19:agu}, but has never been considered for the bow shock).

\section{Conclusions} \label{sec:conclusions}
In this study, we modeled resonant interactions between solar wind electrons and intense short wave-packets of whistler-mode waves. For two magnetic field configurations, corresponding to the foreshock transient and Earth's bow shock, we demonstrate that:
\begin{itemize}
  \item The nonlinear resonant effect, phase trapping, is modified for short wave packets: the probability of trapping becomes larger, whereas energy change due to single trapping becomes smaller. The phase bunching is barely affected by the wave-packet size. Thus, the $\Delta E$-distribution of electron energy change due to a single resonant interaction can be averaged over initial energies to derive the cumulative probability distribution $\mC(\Delta E)=\int_{-\infty}^{\Delta E}{\bar\mP(x)dx}$ for modeling the electron energy evolution $E_{n + 1}  = E_n  + \Delta E\left( {\xi _n } \right)$ with $\mC(\Delta E)=\xi_n$.
  \item We have proposed and verified the probabilistic approach based on $\bar\mP(\Delta E)$ probability distributions to describe electron energy variations.
  \item This probabilistic approach can be used to model the long-term dynamics of electron distributions in a system with multiple nonlinear resonant interactions with short wave-packets.
\end{itemize}
In the second paper, we will further generalize this approach for systems with a finite probability of electron resonant interactions with long wave-packets.

\section*{Acknowledgments}
X.S., A.V.A., X.-J.Z., and V.A. acknowledge THEMIS contract NAS5-02099, NASA grants 80NSSC22K1634, 80NSSC21K0581 (spacecraft data analysis and numerical simulations). A.V.A. and D.S.T. also acknowledge Russian Science Foundation through grant No. 19-12-00313 (theoretical models).

\section*{Data Availability}
This is a theoretical study, and all figures are plotted using numerical solutions of equations in the paper. The data used for figures and findings in this study are available from the corresponding author upon reasonable request. Spacecraft data for the first figure are openly available in THEMIS data repository \url{https://themis.igpp.ucla.edu/}.  Data access and processing were done using SPEDAS V4.1 \cite{Angelopoulos19}.

\bibliographystyle{unsrtnat}

\end{document}